\documentclass[iop,dvipdfmx,letterpaper]{emulateapj}

\usepackage{bm}
\usepackage{float}
\usepackage{amsmath}
\usepackage{mathrsfs}

\usepackage{upgreek}
\usepackage{graphicx}
\usepackage{color}

\usepackage{threeparttable}

\newcommand{\msunyr}{M_{\odot}\:{\rm yr}^{-1}}
\newcommand{\mdot}{\dot{m}}

\newcommand{\msun}{M_{\odot}}
\newcommand{\rsun}{R_{\odot}}

\newcommand{\cloudy}{C{\small LOUDY}}
\newcommand{\HII}{\ion{H}{2}}
\newcommand{\Sigcl}{\Sigma_{\rm cl}}
\newcommand{\gcm}{{\rm g\:cm^{-2}}}

\newcommand{\II}{I\hspace{-.1em}I}
\defcitealias{jon90}{Paper I}

\renewcommand{\thefootnote}{\fnsymbol{footnote}}

\shorttitle{Outflow-Confined \HII~regions. \II. The Early Break-Out Phase}
\shortauthors{Tanaka et al.}

\begin{document}

\title{Outflow-confined \HII~Regions. II. The Early Break-Out Phase}

\author{Kei E. I. Tanaka}
\affil{Department of Astronomy, University of Florida, Gainesville, FL 32611, USA; ktanaka@ufl.edu}
\author{Jonathan C. Tan}
\affil{Departments of Astronomy \& Physics, University of Florida, Gainesville, FL 32611, USA}
\author{Jan E. Staff}
\affil{College of Science and Math, University of Virgin Islands, St. Thomas, United States Virgin Islands 00802, USA}

\and
\author{Yichen Zhang}
\affil{The Institute of Physical and Chemical Research (RIKEN), Hirosawa 2-1, Wako-shi, Saitama, 351-0198, Japan}

\begin{abstract}

In this series of papers, we model the formation and evolution of the
photoionized region and its observational signatures during massive
star formation.  Here we focus on the early break out of the
photoionized region into the outflow cavity.  Using results of 3-D
magnetohydrodynamic-outflow simulations and protostellar evolution
calculations, we perform post-processing radiative-transfer. The
photoionized region first appears at a protostellar mass of
$m_*=10\msun$ in our fiducial model, and is confined to within
$10$--$100\rm\:AU$ by the dense inner outflow, similar to some
observed very small hypercompact \HII~regions.  Since the ionizing
luminosity of the massive protostar increases dramatically as
Kelvin-Helmholz (KH) contraction proceeds, the photoionized region
breaks out to the entire outflow region in $\la10,000\rm\:yr$.
Accordingly, the radio free-free emission brightens significantly in
this stage.  In our fiducial model, the radio luminosity at
$10\rm\:GHz$ changes from $0.1\rm\:mJy~kpc^2$ at $m_*=11\msun$ to
$100\rm\:mJy~kpc^2$ at $m_*=16\msun$, while the infrared luminosity
increases by less than a factor of two. The radio spectral index also
changes in the break-out phase from the optically thick value of
$\sim2$ to the partially optically thin value of $\sim0.6$.
Additionally, we demonstrate that short-timescale variation in
free-free flux would be induced by an accretion burst. The outflow
density is enhanced in the accretion burst phase, which leads to a
smaller ionized region and weaker free-free emission. The radio
luminosity may decrease by one order of magnitude during such bursts,
while the infrared luminosity is much less affected,
since internal protostellar luminosity dominates over accretion
luminosity after KH contraction starts. Such variability may be
observable on timescales as short $10$--$100\rm\:yr$, if accretion
bursts are driven by disk instabilities.
\end{abstract}


\section{Introduction} \label{sec_intro}

Massive stars are the main sources of energetic and chemical feedback
in most galaxies via their radiation, wind, and supernovae. However,
the formation of massive stars is not understood well compared to
low-mass star formation \citep{tan14}. 
One widely discussed class of models is based on the Core Accretion
scenario, e.g., the Turbulent Core model of \citet{mck03}.  These
models are scaled-up versions of the standard scenario of low-mass
star formation \citep{shu87}: massive stars are formed from massive
prestellar cores that are supported largely by nonthermal pressure
forces, i.e., due to turbulence and magnetic fields.
However, there is also significant differences compared to low-mass star formation,
such as the degree of photoionization around the protostar.
Massive stars are expected to start nuclear burning while still
accreting. Their luminosity and photospheric temperature become high
and they then radiate vast amounts of Lyman continnum photons with
energies above $13.6\rm\:eV$, which ionize H to create \HII~regions.
These protostellar \HII~regions emit radio free-free continuum,
providing observational clues of the massive star formation process.

The radio emission from massive protostars has been studied for a long
time \citep[e.g.,][]{mez67,woo89,kur94,hoa07}.  Ultra-compact
($\le0.1\rm\:pc$) and, especially, hyper-compact ($\le0.01\rm\:pc$)
\HII~regions may include sources that are in the protostellar phase.
Elongated radio continuum sources, refereed to as radio jets, have
been seen in a number of sources \citep[e.g.,][]{guz12}.  Recently,
\citet{ros16} performed a high-sensitivity radio continuum survey with
the Karl G. Jansky Very Large Array (VLA) toward early stages of
massive star-forming molecular clumps.  They showed that the radio
detection rates increase from cold infrared dark clouds (IRDCs) to
molecular cores, which suggests the radio emission is a useful tracer
of the evolutionary stages of massive star formation.  Theoretical
models are essential to interpret these observational signatures of
evolution.

Here we present the second of a series of papers on modeling the
photoionization process and its observational signatures of the
Turbulent Core model of \citet{mck03}. In the same context,
\citet{zha14} and \citep{zha17} have developed a suite of
radiative transfer models from infrared (IR) to sub-millimeter
(sub-mm) wavelength \citep[see also][as their paper
  series]{zha11,zha13,zha17}. In our first paper \citep[][hereafter Paper
  I]{KT16}, we calculated the evolutionary sequence of photoionized
structures using the infall-and-outflow structures developed by
\citet{zha14}. We showed that at early stages the photoionized region
is tightly confined by the dense wall of the inner outflow. Later the
photoionized region breaks out to fill the entire outflow cavity in a
timescale of $10^3$--$10^4\rm\:yr$. The predicted free-free flux and
the width of Hydrogen recombination lines from these models are
consistent with the trend of observed radio jets.  However, our model
grid in Paper I has the protostellar-mass resolution of $\Delta
m_*\ge4\:\msun$ which is too coarse to follow the break-out phase of
the photoionized region in detail. The break-out phase is important as
the first signature of the evolution of the photoionized region, i.e.,
the first observational signature that a massive star is actually
forming.  

In this second paper, we thus increase the protostellar-mass
resolution of our gird model to $\Delta m_*=1\:\msun$ around the
break-out phase.  We also upgrade the outflow density structure from
the analytic model by \citet{zha14} to the simulated model based on
the high-resolution three-dimensional magneto-hydrodynamics (MHD) code
by \citet{sta15}. An additional improvement from Paper I is that we
discuss not only the longterm evolution over $10^4\rm\:yr$ of the
photoionized region and its free-free emission, but also their
potential short-timescale variability over $10$--$100\rm\:yr$.  Recent
observations of massive protostars have reported short-timescale
variability observed in radio free-free continuum emission
\citep{car15} and in thermal continuum emission from dust
\citep{car16,hun17}. Also theoretical simulations have suggested
accretion bursts may occur via disk instabilities in massive star
formation \citep{kru09,kui11,mey17,mat17}. The photoionized structure is
sensitive to the outflow density profile, which is expected to scale
with the mass accretion rate, e.g., in magnetocentrifugally-driven
disk wind models. Therefore, we investigate the potential of the
short-timescale variability of the photoionized structure and its
free-free emission as a response to variable accretion rates.

\section{Methods}\label{sec_method}

\begin{figure*}
\begin{center}
\includegraphics[width= 120mm]{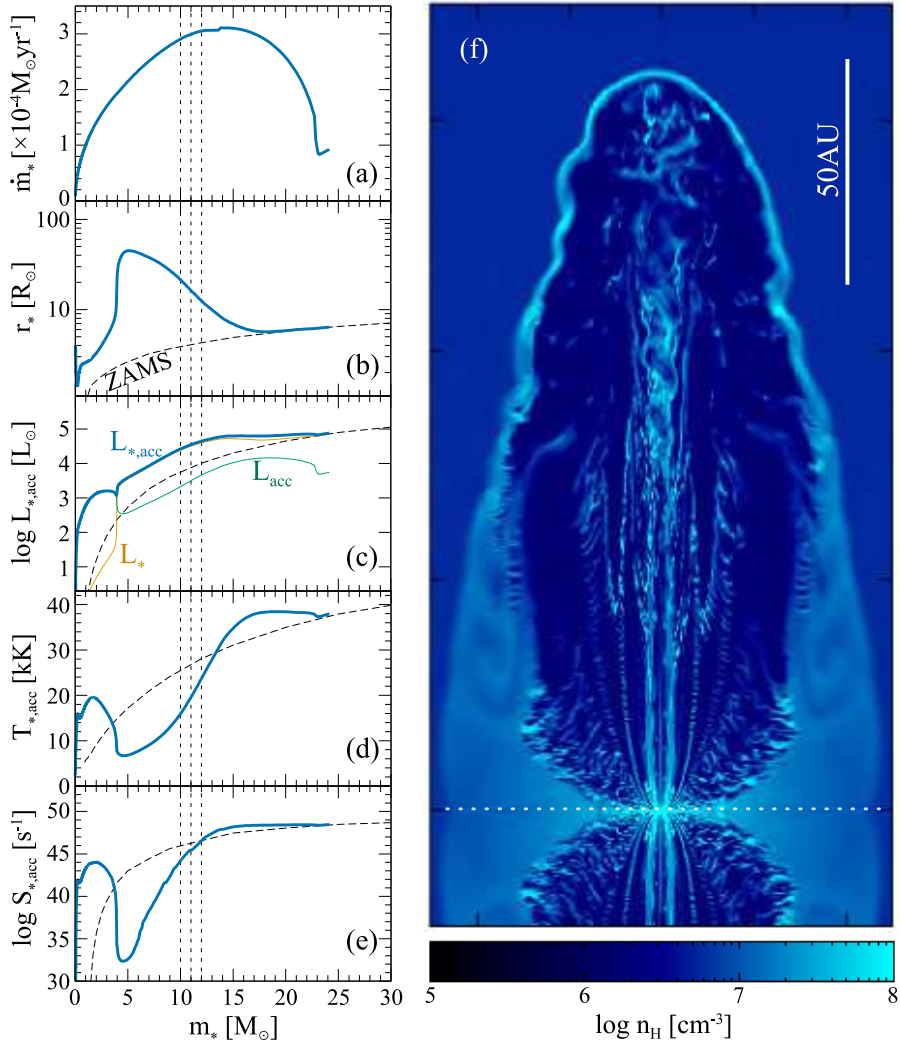}
\end{center}
\caption{
The protostellar and outflow models.  ({\it Left panels a - e}:)
Evolution of the accretion rate ($\mdot_{*}$), protostellar radius
($r_*$), total luminosity including that from accretion ($L_{\rm
  *,acc}$), effective surface temperature ($T_{\rm *,acc}$), and total
H-ionizing photon production rate ($S_{\rm *,acc}$), as functions of
protostellar mass.  The properties of the zero-age main sequence
(ZAMS) are shown by dashed lines \citep[Table 1 of][]{dav11}.
In panel (c), the accretion and intrinsic luminosities ($L_{\rm
acc}$ and $L_*$) are also shown by thin green and orange lines, respectively.
The vertical dotted lines indicate the stages at
$m_*=10$, $11$, and $12\:\msun$, which are inputs for post-processing
photoionization calculations.  The protostellar evolution is
calculated by the method of \citet{KT17} (see text).
({\it Right panel f}:) Outflow density structure (in a thin slice) at the end of
the simulation by \citet{sta15}.
Here the density structure at a protostellar mass of $12\:\msun$ is shown, however the structure is
also scalable to other cases (see \S\ref{sec_MHD}).  Mirror
symmetry is assumed about the disk mid-plane, which is shown by the
white dotted line.}
\label{fig_method}
\end{figure*}

\subsection{Protostellar models} \label{sec_prost}

We calculate protostellar evolution based on the semi-analytic method
of \citet{KT17}, which self-consistently included mass accretion
regulated by multiple feedback processes \citep[see also the series of
  papers by][]{zha11,zha13,zha14}. 
This model assumes a prestellar core collapses to form one massive
star.
The initial cloud core is assumed to be spherical and in quasi-virial
equilibrium, supported by turbulence and/or magnetic fields
\citep{mck03}. The core properties are determined by three main
parameters; core mass $M_c$, mass surface density of the ambient clump
$\Sigcl$, and the core's initial rotational to gravitational energy ratio $\beta_c$.
In this paper, we set fiducial core mass and surface density to
$M_c=60\:\msun$ and $\Sigcl=1\:\gcm$ \citep{mck03}.
The core radius is then determined to be $0.057{\rm\:pc}$.
Based on observations of both low-mass and high-mass cores,
$\beta_c$ is found to have a typical value of $\sim0.02$ \citep[e.g.,][]{goo93,li12,pal13}.
Therefore, we fix the fiducial value of 0.02 for $\beta_c$.

The cloud core undergoes the inside-out collapse which is treated by a
self-similar solution \citep{mcl97}. The accretion rate onto the
central star $\mdot_*$ is shown in panel (a) of Figure \ref{fig_method}.
The typical value of $\mdot_*$ is $\sim10^{-4}\:\msunyr$.
The accretion is regulated by multiple feedback processes, i.e.,
MHD disk winds, radiation pressure, photoevaporation, and stellar winds
\citep{KT17}.  Eventually, the main accretion phase finishes at a
stellar mass of $24\msun$, i.e., the star formation efficiency from
the initial core is $24\msun/60\msun=0.4$.

The protostellar evolution is obtained based on the model of
\citet{hos09} and \citet{hos10}. 
This model solves for the internal structure of
the protostar, assumed to be spherical, including effects such as
hydrogen and deuterium burning, convection, and radiative transfer. 
The evolution of the protostar depends on the accretion
history, which is self-consistently calculated as explained above.
The evolution of the accretion rate and protostellar
properties are shown in panels (b) - (e) in Figure \ref{fig_method}.
The stellar radius, $r_*$, first increases to be as large as
$45\:\rsun$ when $m_*=5\msun$, due to the relatively high
mass-accretion rate. Then the protostar undergoes Kelvin-Helmholtz
(KH) contraction, reaching the zero-age main sequence (ZAMS) phase at
around $20\:\msun$.  The accretion luminosity, $L_{\rm acc}$,
dominates before KH contraction sets in, while the internal stellar
luminosity, $L_*$, becomes dominant later. The total luminosity,
$L_{\rm *,acc}$, increases almost monotonically. The effective
temperature, $T_{\rm *,acc}$, has a local minimum when $m_*=5\:\msun$
because of the large stellar radius. By the combination of variations
of $L_{\rm *,acc}$ and $T_{\rm *,acc}$, the ionizing-photon rate,
$S_{\rm *,acc}$, drops down to $10^{33}\:{\rm s^{-1}}$ at around
$5\msun$.  It then rises dramatically to over $10^{48}\:{\rm s^{-1}}$
during KH contraction.  This leads to the formation and expansion of a
photoionized region in the KH contraction phase (Paper I). To study
the earliest phase of photoionization, the protostellar models in the
KH phase at stellar masses of $m_*=10$, $11$, and $12\:\msun$ are
adopted for the following photoionization calculations, which are
named as models M10, M11, and M12, respectively.  The time duration
from $m_*=10$ to $12\msun$ is $6,800{\rm\:yr}$ in this evolutionary
calculation.

\subsection{Outflow density profile}\label{sec_MHD}

The outflow density profile is needed to calculate the photoionized
structure.  In Paper I, we adopted the semi-analytic disk-wind model,
which is modified from the magnetocentrifugally-driven outflow model
of \citet{bla82} \citep{zha13}.  The mass-loading rate of the
wind is proportional to the mass accretion rate with a ratio $f_{\rm
  w}=0.1$ which is a typical value for disk winds \citep{kon00}.
Though our analytic disk-wind model appears to have some agreement with observed
sources \citep[Paper I]{zha13a}, it is highly simplified and
idealized, e.g., assuming an axisymmetric smooth structure.  In this
study, therefore, we apply the density profile from the
three-dimensional MHD simulation of disk winds by \citet{sta15} for a
post-processing photoionization calculation.

\subsubsection{Models for fiducial accretion rate}\label{sec_stand}

\citet{sta15} studied disk winds from a disk using the ZEUSMP code
\citep{nor00}.
This is a three-dimensional
ideal MHD simulation with a Keplerian accretion disk as a fixed
boundary condition, along with a magnetic field that is initially
poloidal and in a self-similar configuration.
The wind launched from the disk surface is magneto-centrifugally accelerated
and thus creates the outflow.
Although the original goal of \citet{sta15} was the simulation of jets from low-mass protostars,
this simulation is also fully scalable to the outflows around massive protostars.
This is because they solved the dimensionless form of the equations,
and thus the results can be re-scaled based on following fundamental protostar properties
\citep[more details please see][]{ouy97}.
The unit length (and also the highest resolution)
is set by the the innermost radius of the accretion disk.
The innermost radius may be larger than the stellar radius if the
stellar magnetosphere is strong enough to truncate the disk
\citep{gho78,shu00}.
However, in the mass-accretion phase which we are interested in,
the accretion disk may continue to the surface of the star,
because the kinematic and gaseous pressures are higher than the magnetic pressure due to the high accretion rate.
Therefore, we chose the disk innermost-radius as the stellar radius, $r_*$.
The time unit is the inverse of the Keplerian angular velocity
at the length unit $r_*$, i.e., $1/\Omega_{\rm K*}\simeq0.18\left(
m_*/10\msun \right)^{-1/2} \left(r_*/10\rsun \right)^{3/2}$.  The
density unit is set to match the simulated disk-wind rate to the
analytic value of $f_{\rm w}\mdot_*$.

While \citet{sta15} have investigated several initial magnetic field
configurations, we use the result of the ``BP'' configuration, which
is consistent to the model of \citet{bla82}, i.e., $B_{\rm p}\propto
\varpi^{-1.25}$, where $B_{\rm p}$ is the initial poloidal magnetic
field strength at the disk surface and $\varpi$ is a cylindrical
radius. Panel (f) of Figure~\ref{fig_method} shows a density slice for
the outflow from our $m_*=12\:\msun$ protostar model. Mirror symmetry
at the disk mid-plane is assumed. The density profile is quite close
to being axisymmetric, but with smaller non-axisymmetric structures,
which could not be treated in the previous semi-analytic model we used
in Paper I. Note that the geometric thickness of the disk is assumed
to be zero in this simulation,
so the model applies to the region of outflow cavity away from the disk region.

We note a caveat on the usage of this simulated outflow. As a result
of its high spatial resolution, the simulated time is limited to be
shorter than the expected accretion time. Therefore, the size of the
simulated outflow is only about $100{\rm AU}$ (panel (f) in
Fig. \ref{fig_method}), while, in reality, the outflow is expected to
be larger than the core scale of $0.1\:{\rm\:pc}$. On the other hand,
thanks to this high resolution, we are able to model structures down
to the vicinity of the protostellar radius where the stellar radiation
is mainly processed.
Thus, in this paper, we focus only on the earliest break-out phase of the photoionized region while it is
confined on scales of $\sim100{\rm\:AU}$ or smaller.
In this earliest phase,
the rest of the outer outflow would be essentially neutral and would not contribute significantly to the free-free emission due to photoionization.

\subsubsection{Models for variable accretion bursts}

In this paper, we also address the potential short-timescale
variability induced by accretion bursts. The accretion rate in our
evolutionary model described in \S\ref{sec_prost} changes smoothly on
a timescale of $10^{4}$--$10^{5}{\rm\:yr}$.  However, some numerical
simulations of massive star formation predict that the accretion rate
can be highly time-variable due to disk self-gravity instabilities
\citep{kru09,kui11,mey17,mat17}. In the episodic accretion scenario, the
protostars spend most of their time in a low accretion-rate, quiescent
phase, with short-timescale ($\sim$tens of years) accretion bursts in
which accretion rates are one or two orders of magnitude
higher\citep{vor15}.

The outflow is powered by accretion and thus its density is enhanced
in the accretion burst phase \citep[e.g.,][]{tom17,mat17}.  Especially, on
the small scales of $100{\rm\:AU}$, the outflow density would
quickly respond to the accretion variation since the crossing
timescale of outflow is only $\la10{\rm\:yr}$.
{Here we have used the outflow velocity of
 $\sim250$--$600{\rm\:km\:s^{-1}}$ from the MHD simulation of \citet{sta15}.

To investigate the variation of the photoionized structure induced by
episodic accretion, we mimic the burst/quiescent phases by
enhancing/reducing the outflow-density by a factor of 3.16 from the
standard value for the $11\msun$ model, which we refer to as M11H and
M11L, respectively.
The density variation between models M11H and
M11L is one order of magnitude, i.e., $3.16^2\simeq10$. Note that
the accretion bursts may in general induce density changes that
are larger or smaller than this factor. In such cases, the
resulting variation in radio flux would be greater or smaller than
those shown in \S\ref{sec_burst}, respectively.
We use the same protostellar model as the standard accretion case, although the
episodic accretion may also affect the protostellar properties.  We
discuss this point further in \S\ref{sec_discussion}.

\subsection{Photoionizing calculation} \label{sec_phion}

We calculate the photoionized structure using the protostellar
properties from the stellar evolution calculation (\S\ref{sec_prost})
and the outflow density profile from the MHD simulation
(\S\ref{sec_MHD}). In Paper I, under the assumption of axisymmetry,
the full transfer of ionizing photons was solved, which allows the
accurate treatment of both the direct and diffuse radiation fields.
As the density profile from the 3-D simulation is used in this study,
it is computationally much more expensive to perform the full
radiative transfer calculation. Thus we simplify the method of the
photoionizing calculation. \citet{KT13} showed that the direct stellar
radiation is more important than the diffuse radiation from
recombination to the ground state.  In this study, we use the
on-the-spot approximation to treat the diffuse radiation field, i.e.,
the diffuse radiation is assumed to be absorbed at the point of
emission. Then, the photoionized radiation field can be obtained by
\begin{eqnarray}
\frac{1}{r^{2}} \frac{d}{dr} \left(r^2 F_{\rm EUV}\right)  =-(1-x_{\rm II})n_{\rm H} \sigma_{\rm H} F_{\rm EUV},\label{eq1}
\end{eqnarray}
where $F_{\rm EUV}$ is the radial number flux of ionizing photons, $r$
is the distance from the center of the star, $n_{\rm H}$ is the number
density of hydrogen nuclei, $x_{\rm II}$ is the ionization fraction of
H, and $\sigma_{\rm H}$ is the mean photoionization cross-section of
the hydrogen atom. The transfer equation (\ref{eq1}) can be solved as,
\begin{eqnarray}
F_{\rm EUV}=\frac{S_{\rm *,acc} e^{-\uptau_{\rm EUV}}}{4\pi r^2}, \label{eq_flux}\\
\uptau_{\rm EUV}(r) = \int_{r_*}^{r} (1-x_{\rm II}) n_{\rm H} \sigma_{\rm H} dr',
\end{eqnarray}
where $\uptau_{\rm EUV}$ is the optical depth for the ionizing
radiation.  The ionization fraction $x_{\rm II}$ is obtained from the
balance between photoionization and recombination,
\begin{eqnarray}
(1-x_{\rm II})n_{\rm H} \sigma_{\rm H} F_{\rm EUV}=\alpha_{\rm B}(T) x_{\rm II}^2 n_{\rm H}^2, \label{eq_balance}
\end{eqnarray}
where $\alpha_{\rm B}(T)$ is the recombination coefficient to all
states except the ground state (so-called case B) which depends on the
local gas temperature $T$.  The temperature of photoionized gas is
usually regulated around $10,000{\rm\:K}$.  We set the temperature as
a function of density and the stellar spectrum, based on parameterized
spherical calculations using \cloudy\:\citep{fer13} (see \S2.2.3 of
Paper I).  The mean values of the hydrogen cross-section, $\sigma_{\rm
  H}$, is evaluated from the stellar spectra for each model. Solving
equations (\ref{eq_flux}) -- (\ref{eq_balance}) over the entire solid
angle, the photoionized structure is obtained.  In this paper, the
dust scattering and absorption of ionizing photons are ignored which
was included in Paper I. We expect this approximation to be valid,
since here we focus on the innermost region of $<100\rm\:AU$, where
the dust grains have been destroyed, i.e., if most of the volume is
filled by outflows launched from the region of the disk inside the
dust destruction front.

\subsection{Free-free emission}\label{sec_ff}

We calculate the free-free radio continuum emission from the obtained
photoionized outflow structures. The method for the free-free transfer
calculation is the same as Paper I, but now applied to the 3-D calculation
of this study. The equation for the free-free radiation transfer is,
\begin{eqnarray}
\frac{dI_{\nu,{\rm ff}}}{ds}=-\kappa_{\nu,{\rm ff}}
\left( I_{\nu,{\rm ff}} - B_{\nu} \right)
\end{eqnarray}
$I_{\nu,{\rm ff}}$ is the free-free intensity, $\kappa_{\nu,{\rm ff}}(T, n_{\rm H})$ is the opacity of free-free emission, and
$B_{\nu}(T)$ is the Planck function.  The observed flux density is
calculated by integrating over source solid angle of $\Omega_{s}$,
\begin{eqnarray}
F_{\nu, {\rm ff}}
= \int_{\Omega_{s}} I_{\nu, {\rm ff}} (\theta_{\rm view}) d\Omega.
\end{eqnarray}
Here the far field limit is assumed.
See also \S2.3.1 of Paper I for more details of the radiative transfer
of free-free emission.

\begin{figure}
\begin{center}
\includegraphics[width= 90mm]{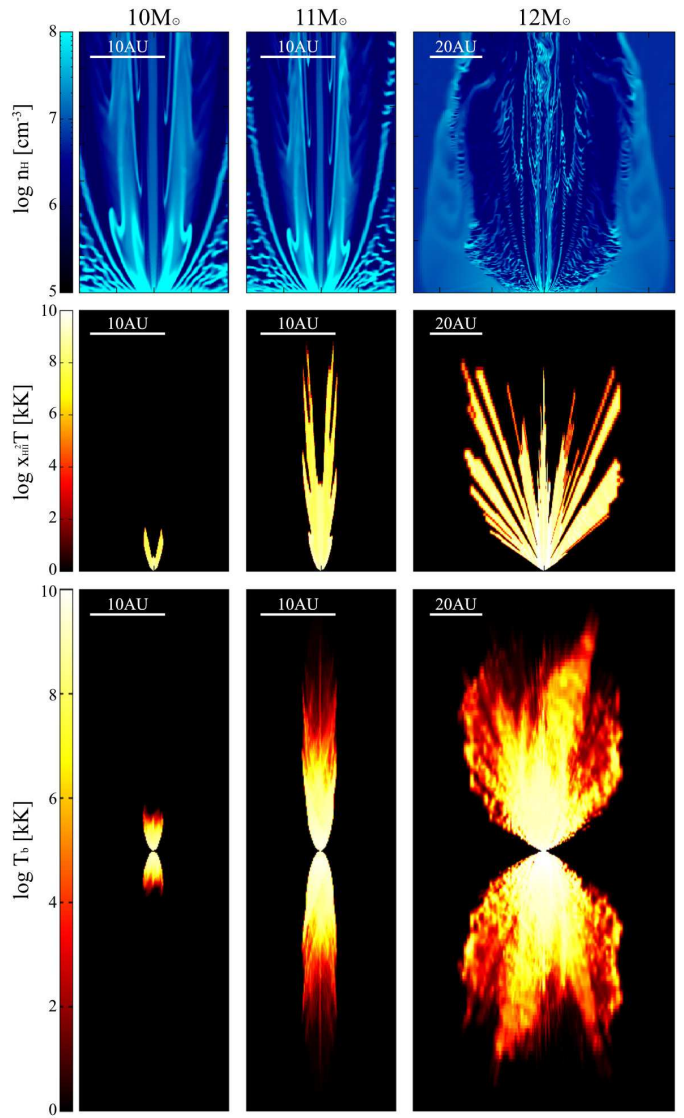}
\end{center}
\caption{
The earliest evolution of the outflow-confined \HII~region for models
M10, M11, and M12 (from left to right), shown by density slices (top),
ionized-fraction-weighted temperature slices (middle), and the
resolved free-free emission images at $10{\rm\:GHz}$ (bottom).  The
scale-length is shown in each panel.  In the bottom panels, the
intensity is shown by the brightness temperature and a viewing angle
of $60\:\degr$ is assumed.}
\label{fig_HII1}
\end{figure}

\section{Results} \label{sec_results}

We here describe the results of photoionizing and free-free emission
calculations.
First, we show the earliest break-out evolution of the photoionized region
which occurs over a time of about $10^4{\rm\:yr}$. 
Then, we explain the variation of the photoionized structure due to
accretion bursts, which could occur in tens of years.

\subsection{Earliest break-out of Outflow-Confined \HII~Regions}

The evolution from $m_*=10\msun$ to $12\msun$ is shown in
Figure~\ref{fig_HII1} (models M10, M11, and M12).  In the top panels,
the outflow densities of different models are similar, since they are
simply scaled based on the protostellar parameters as described
\S\ref{sec_MHD}. In the middle panels, the displayed temperature
structure is weighted by the square of the ionization fraction, i.e.,
$x_{\rm II}^2 T$, because we are only interested in the photoionized
region and do not solve for the temperature of the neutral region in
this study.  Recall that, in the photoionization calculation, most of
the region is either fully-ionized ($1-x_{\rm II}\ll1$) or neutral
($x_{\rm II}\ll1$). The photoionized region, where the temperature is
about $10,000{\rm\:K}$, is confined on $100{\rm\:AU}$ scales by the
high-density outflow, i.e., an ``outflow-confined'' \HII~region
\citep{tan03}. Especially, the size of the photoionized region is only
about $10{\rm\:AU}$ when $m_*=10\:\msun$. Then, the photoionized
region elongates near the outflow axis at $11\:\msun$, since the
density is lower in this direction than that near the equatorial
plane. The photoionized region starts to expand widely at $12\:\msun$.
The shape of the photoionized region is very spiky. This is because of
the shadows cast by the high-density clumpy structures.

The bottom panels of Figure \ref{fig_HII1} show the free-free emission
maps of models M10, M11, and M12 at $10\rm\:GHz$.  The viewing angle
from the outflow axis is assumed to be $60\degr$.  The spiky features
in the temperature slices are now less apparent because of their
overlap along the line of sight.  However, the bipolar shape and the
elongated structure are still visible, especially in model M11.

\begin{figure*}
\begin{center}
\includegraphics[width= 170mm]{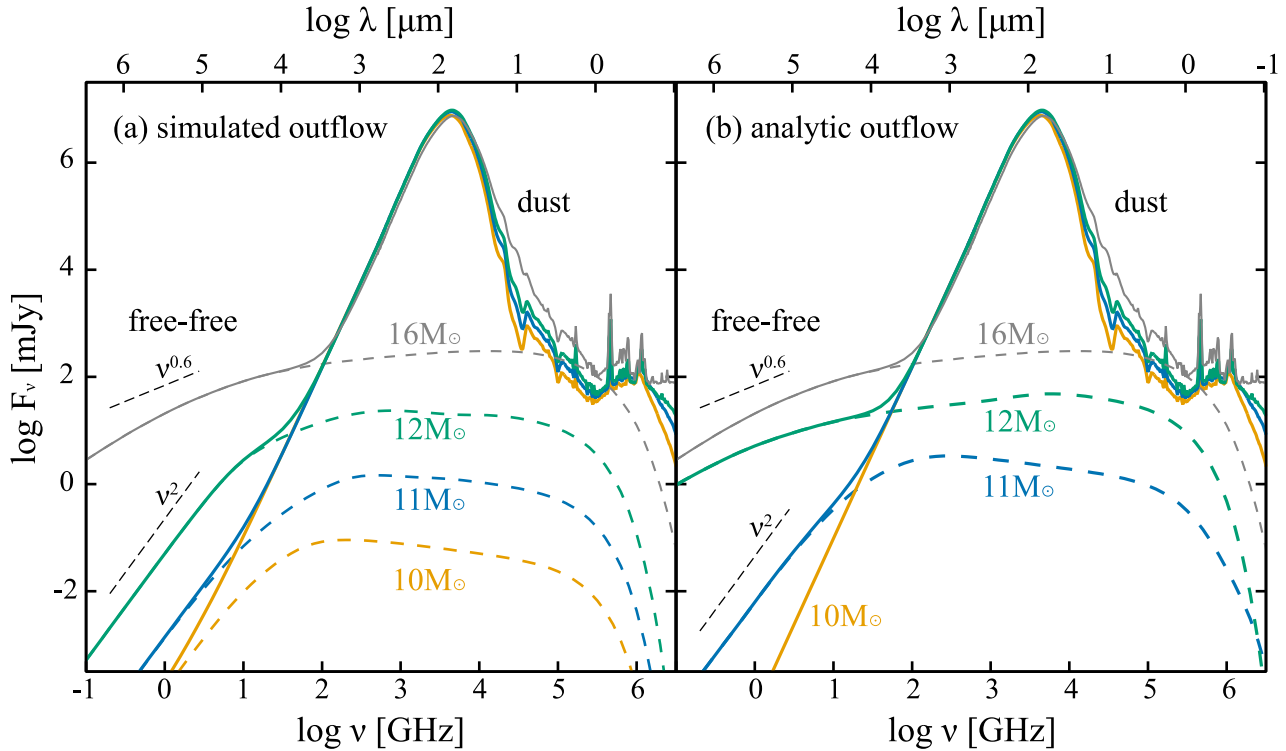}
\end{center}
\caption{
(a) SEDs for models M10, M11, and M12, together with the case of
  $m_*=16\msun$ from Paper I
(orange, blue, green, and gray, respectively).
(b) For comparison, SEDs for $m_*=10,~11,~12$, and $~16\:\msun$ based
on the analytic outflow density model calculated as in Paper I
(orange, blue, green, and gray, respectively).  In both
panels, the thick dashed lines represent free-free emission, while
the thick solid lines show the total emission including dust
emission.  The black thin-dashed lines are plotted for
illustrative purposes, to show spectral indices of 0.6 and 2.  Note
that the dust thermal emission is obtained from \citet{zha17}, and a
distance of $1\rm\:kpc$ and the viewing angle of $60\degr$ are assumed
in all models. The free-free emission for the $10\msun$ model in panel
(b) is not seen because it is weaker than $10^{-3}\rm\:mJy$.}
\label{fig_SED1}
\end{figure*}

Figure \ref{fig_SED1}(a) shows the SEDs for models M10, M11, and M12.
A distance of $1\rm\:kpc$ and a viewing angle of $60\:\degr$ are
assumed. The dust thermal emission is evaluated based on the model of
\citet{zha14}, which typically has a spectrum index of $3.3$ at radio
frequencies. For comparison, the SED of a later stage model with
$m_*=16\:\msun$ from Paper I is also shown in Figure \ref{fig_SED1}(a)
(model A16 in Paper I). Combining our new results of M10, M11, and M12
and the previous $16\msun$ model, the free-free emission evolves in
the following manner. At the moment when the photoionized region is
formed (M10), the free-free flux is lower than the dust flux at
  most frequencies.  Then, the ionized region expands to $100\rm\:AU$
scales in about $10^4\rm\:yr$, and the radio free-free flux starts to
dominate the dust flux at around $10\rm\:GHz$ or lower
  frequency.  (M11 and M12).  At this stage, the free-free flux has a
spectral index of about $2$.  Later, the photoionized region breaks
out beyond the $1000{\rm\:AU}$ scale, most of the outflow cavity is
ionized, and the free-free flux density becomes as high as
$100\rm\:mJy$ at $10\rm\:GHz$ with a spectral index of $0.6$ (models
of Paper I).

While the infrared emission at the peak of the SED increases by
a factor of two from the case with a protostellar mass of $10\:\msun$
to that of $16\:\msun$, the flux at $10\rm\:GHz$ jumps by three orders of magnitude
because of the sharp rise of the ionizing photon rate due to the KH
contraction.  The spectral index of near $2$ in the earlier stage
indicates the ionized region is mostly optically thick at these
wavelengths. This is because the density of the inner
$100\rm\:AU$-region is very high.  On the other hand, the shallower
index of $\sim0.6$ in the later stage represents partially optically
thin conditions, since the outer region has lower density \citep[for
  more details about the partially optically thin case, see \S4.1 of
  Paper I and a classic model by][]{rey86}.

In Figure \ref{fig_SED1}(b), for comparison, we present the SEDs
calculated using the analytic outflow density structures from Paper I.
The basic evolution of free-free emission is the same as that in the
simulated outflow case: the flux rises up by orders of magnitude and
the spectral index changes from $\sim2$ to $\sim0.6$ as the
protostellar mass increases from $m_*=10$ to $16\msun$.  The main
difference in free-free evolution of the simulated and analytic
outflow models is the degree of increase from $10$ to $12\msun$ at
radio wavelength.  In the case of the simulated outflow model, the
free-free flux changes about two orders of magnitude from $m_*=10$ to
$12\:\msun$ at $\sim10{\rm\:GHz}$.  On the other hand, in the case of
the analytic outflow model, the free-free flux is weaker than
$\sim{\rm \mu Jy}$ at $10\msun$ and then increases to the $\sim{\rm
  mJy}$ level at $12\:\msun$.  This is because the analytic outflow
has a steeper density gradient near the outflow axis than the
simulated outflow and thus the photoionized region breaks out more
quickly.
The difference of break-out timing also
causes differences in the spectral indices at $\nu<10{\rm\:GHz}$
in the case of $m_*=12\msun$: the index is close to 2 in the
analytic model since the photoionized region is confined within a
$100\rm\:AU$ scale, where it is optically thick at this frequency,
while the index of the simulated model is close to 0.6 because the
photoionized region is extended to include optically thinner outer
regions.
In this way, the details of photoionized-region
break-out are sensitive to the outflow structure, while the general
trends of the evolution are not.

\subsection{Variations Induced by Accretion Bursts} \label{sec_burst}

Next, we examine the potential variability induced by accretion
bursts. Model M11L represents the quiescent phase which has lower
outflow density by a factor of 0.316 than the standard case of M11,
and model M11H represents the burst phase having the higher density by
a factor of 3.16 than the standard case.

\begin{figure*}
\begin{center}
\includegraphics[width= 170mm]{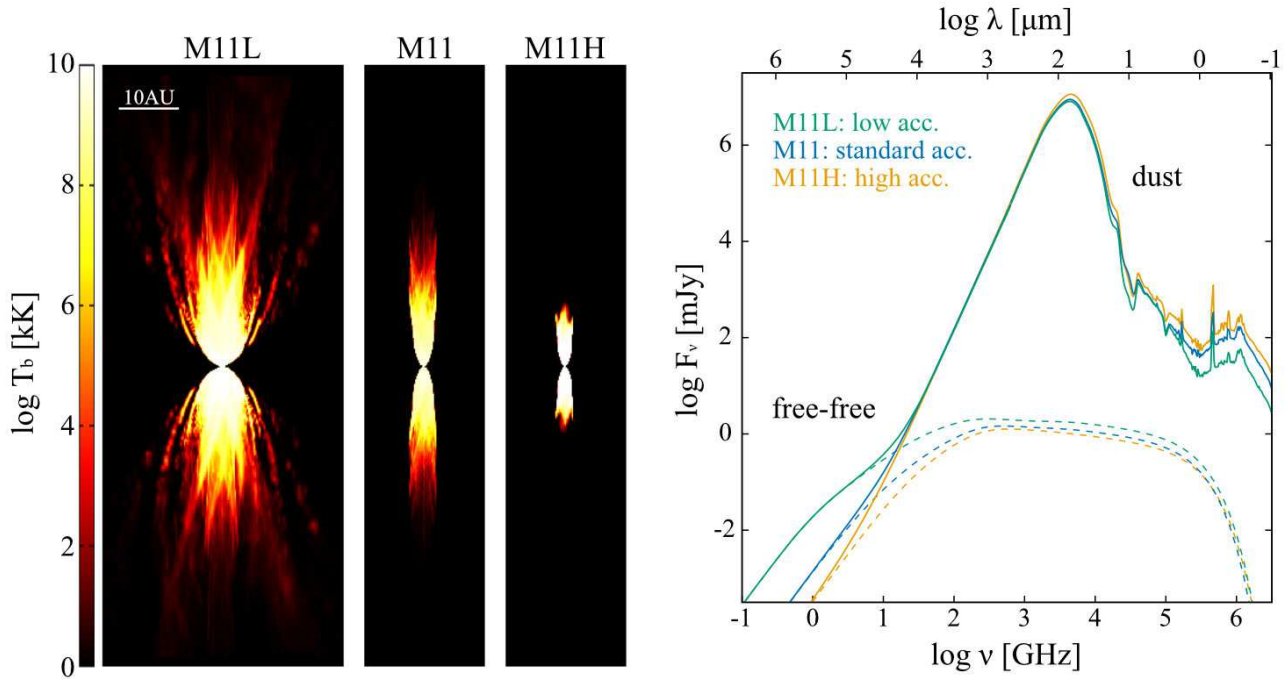}
\end{center}
\caption{
Radio emission variabilty induced by accretion bursts. {\it (Left:)}
Resolved free-free images for models M11L, M11, and M11H (left to
right). The intensity is shown as brightness temperature for emission
at a frequency of $10\rm\:GHz$. The linear scale is same in all
panels, with sources assumed to be at 1~kpc distance and a viewing
angle of $60\degr$ from the outflow axis.
{\it (Right:)} SEDs for models M11L, M11, and M11H (orange, blue,
  green, and gray, respectively), again for a distance of $1\rm\:kpc$
and viewing angle of $60\:\degr$ to the outflow axis. The dashed
  lines show free-free emission, and the solid lines represent the
  total emission including dust emission. 
  The thermal dust emission is obtained from \citet{zha17}.
Note that the radio flux at frequencies lower than $10\rm\:GHz$
decreases by an order of magnitude from a higher value in the low
accretion phase to a lower value during the accretion burst, while the
peak of the thermal dust emission changes by a much smaller factor.}
\label{fig_burst}
\end{figure*}

The left panels in Figure~\ref{fig_burst} show the radio free-free
continuum images of M11L, M11, and M11H at $10{\rm\:GHz}$.  In the
quiescent model M11L, the photoionized region extends to larger
distances than the standard model M11 because the outflow density is
lower.  However, due to the low density, the outer part of the
photoionized region becomes optically thin and the free-free emission
from there is fainter than the optically thick value of $T_{\rm
  b}\simeq10^4\rm\:K$.  On the other hand, in the accretion-burst
model M11H, the photoionized region is confined to within
$\sim10\rm\:AU$ by the high density outflow and is optically thick at
this frequency.  In this way, the quiescent phase has a larger size
and partially optically-thin emission, while the burst phase shows
smaller and optically-thick emission.

These features are also seen in the SEDs shown in the right panel of
Figure~\ref{fig_burst}. Model M11L has about one order of magnitude
higher free-free flux at $10\rm\:GHz$ than model M11H, because it has
a larger photoionized region.  On the other hand, the spectral index
at $10\rm\:GHz$ of M11L is $\sim1$, indicating it is partially
optically thick, and is shallower than M11H's index of 2. The
free-free flux from outflow-confined \HII~regions is thus sensitive to
the time-variability of the outflow density.
Also note that since the internal luminosity from the protostar
dominates over accretion luminosity after KH contraction starts, the
flux from thermal dust emission does not change as significantly due
to such accretion bursts.

\section{Discussions and Conclusions} \label{sec_discussion}

Here we discuss the observational predictions based on our results of
the earliest break-out phase of outflow-confined \HII~regions and
their time-variability due to variable accretion.

\subsection{Break-out phase of photoionized regions}

\begin{figure*}
\begin{center}
\includegraphics[width= 170mm]{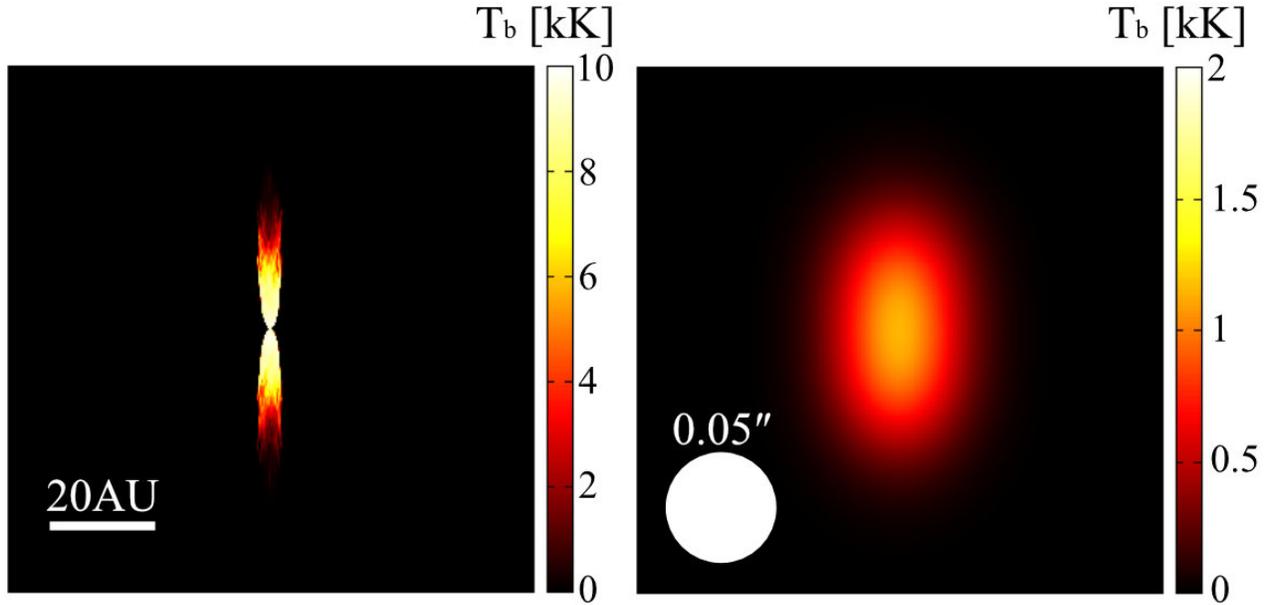}
\end{center}
\caption{
Resolved (left) and beam-convolved (right) images of M11 at
$45{\rm\:GHz}$. In the convolved image, a distance of $500\rm\:pc$ and
beam size of $0.5\:\arcsec$ are assumed. The convolved image has lower
brightness temperature of $\la2,000\rm\:K$, since it is unresolved in
the direction perpendicular to the outflow axis.}
\label{fig_lowTb}
\end{figure*}

The photoionized region starts to form as KH contraction proceeds.
In the model of this paper, the onset of
photoevaporation is at a stellar mass of about $11\:\msun$
(\S\ref{sec_results}).  However, we note that the onset mass
depends on the accretion history because a higher accretion rate
leads to a higher mass at which KH contraction starts, and also
causes the density of outflow to be higher \citep{hos09,zha14}.
For typical accretion rates of $\sim10^{-4}$--$10^{-3}\:\msunyr$,
the mass of the onset of photoionization is $\sim10$--$20\msun$
(see Paper I).
The photoionized region is first confined by the
dense MHD outflow wall at $<100{\rm\:AU}$ scales.  At this stage, the
typical radio luminosity is $\sim{\rm\:mJy\:kpc^2}$, with a spectral
index of $\sim2$, i.e., optically thick conditions.
  \citet{ill16} reported a candidate of this type of hypercompact
  \HII~region around the young massive star G11.92$-$0.61 MM1.
The radio flux increases as the photoionized region
expands, and almost the entire outflow is ionized about
$10^4\rm\:yr$ after its formation. At this stage, the free-free
luminosity becomes $\sim100\rm\:mJy\:kpc^2$, with a spectral index
of $\sim0.6$ at around $10\rm\:GHz$
which is similar to observed
radio jets around massive protostars, e.g., the central radio source
of IRAS16547$-$4247 \citep{rod08}.
See Paper I for more discussion about this later stage.

Considering the highest resolution of the Very Large Array (VLA) of
$0.2\arcsec$ at $10{\rm\:GHz}$ or $0.043\arcsec$ at $45{\rm\:GHz}$ and
typical distances to massive protostars of several $\rm kpc$, it is
usually not possible to observationally resolve the photoionized
region when it is confined within $\la100\rm\:AU$ scales. However,
this becomes possible for sources closer than $1\rm\:kpc$, like Source
I in the Orion KL region, which is $0.4\rm\:kpc$ away \citep{men07}.
In Figure \ref{fig_lowTb}, we show the image of M11 at $45\rm\:GHz$,
convolving by a beam of $0.05\:\arcsec$ FWHM diameter, and assuming a
distance of $0.5\rm\:kpc$. While elongation along the outflow axis can
be resolved, in the perpendicular direction the source is unresolved
since the photoionized region has a very long, thin shape.  This
causes the peak brightness temperature to be $\la2,000\rm\:K$, which
is lower than the ionized gas temperature of $\sim10,000\rm\:K$, even
though it is almost completely optically thick.  Such a model may be
considered as an alternative explanation for the low brightness
temperature, $\sim1,500\rm\:K$, of Source I in Orion KL, rather than
being due to $\rm H^-$ free-free emission from a collisionally ionized
disk \citep{rei07, hir15}.
However, we note
that, while the spectral index of Source I is $1.86\pm0.26$ which
is consistent with our model, its radio luminosity of
$0.2\rm\:mJy\:kpc^2$ is smaller than our model prediction
\citep{for16}.  This low-radio luminosity with the optically-thick
spectral-index might be explained by the outflow-confined
\HII~model with a lower accretion rate.
However, we have studied only one initial condition in this paper.
For more detailed discussion of this photoionized outflow hypothesis
to explain Source I including non-detection of hydrogen recombination lines
at the frequencies of $353.6$ and $662.4\rm\:GHz$ \citep{pla16},
we need to expand the range of protostellar source model
parameters, which we defer to a future paper.

\subsection{Short-time variabilities induced by accretion variation}

Apart from the break-out on $\sim10^3\rm\:yr$ timescales, the
photoionized structure can also change on shorter timescales due to
accretion rate variability. Accretion bursts on $\sim10$--$100\rm\:yr$
timesscales induced by disk instability may be expected during massive
star formation \citep[e.g.,][]{kru09,kui11,mey17,mat17}. Since the
outflow is powered by accretion, the outflow density is higher (lower)
in the accretion burst (quiescent) phase
\cite[see simulations by][]{tom17,mat17}.
Due to this density variation, the photoionized region is smaller in the burst phase than that in the quiescent phase
(\S\ref{sec_burst}).
However, there are as yet no clear observational examples that fully support this theoretical expectation.
\citet{car15} reported that, around the massive protostar W75N(B)-VLA 2,
the radio continuum emission changed from a compact to an elongated morphology within a period of 18 years.
Although the timescale is consistent with those that may result from accretion and outflow variability,
other observational aspects are not particularly supportive of this being an example of photoionized-structure variation induced by accretion/outflow fluctuations.
In particular, water masers, which are expected to trace shocked regions, also change their morphology systematically with the radio continuum, potentially indicating that this radio flux is induced from shocked ionization rather than photoionization.
However, following the recently reported dust-reprocessed luminosity bursts around massive protostars \citep{car16,hun17},
which are expected to be due to accretion bursts,
we predict that the radio flux variations induced by the accretion/outflow fluctuations affecting inner photoionized regions will eventually be observed.

Note that the free-free flux from the photoionized region is smaller
in the accretion burst phase, which is the opposite behavior to that
of the dust luminosity. Additionally our model shows that, when the
accretion rate changes by one order of magnitude, the free-free
emission also varies by one order of magnitude, while the peak of the
thermal dust emission varies much less (Fig.~\ref{fig_burst}).  This
is because the total luminosity is dominated by the internal
protostellar luminosity after KH contraction starts (see
Fig.~\ref{fig_method}).  The accretion luminosity is of greater
relative importance in the earlier protostellar phase.  Therefore,
massive protostars associated with the dust-emitted luminosity bursts
\citep{car16,hun17} may be more likely to be in the pre-KH contraction
phase. We thus suggest that radio free-free observations will be a
better method for tracing accretion variability in the later, more
evolved stages.

Finally, we note the caveat that we used the same protostellar
parameters for the different accretion rate models in
\S\ref{sec_burst}, although protostellar evolution is expected to be
related to accretion history.
\citet{smi12} showed that, in the case of primordial star formation,
protostellar evolution under the influence of a variable accretion
rate is consistent with the evolution calculated with accretion rates
averaged over the last $100\rm\:yr$. This is because the timescale of
protostellar evolution is described by the averaged accretion rate
over the local KH time, $t_{\rm KH}=Gm_*^2/(r_*L_{\rm*acc})$. 
The KH timescale of the protostar considered in this study is about
$5,000\rm\:yr$, which is longer than the typical duration of accretion
bursts, and thus we expect the accretion burst does not alter these
protostellar properties significantly.

\acknowledgements We thank Viviana Rosero, Kazuhito Motogi,
and Taishi Nakamoto for fruitful discussions and comments.
J.C.T. acknowledges support from NSF grants AST 1212089 and AST 1411527.
J.E.S acknowledges support from NASA grant NNX15AP95A.


\clearpage


\begin{thebibliography}{}

\bibitem[Blandford \& Payne(1982)]{bla82}
	Blandford, R.~D., \& Payne, D.~G.\ 1982, \mnras, 199, 883 

\bibitem[Caratti o Garatti et al.(2016)]{car16}
	Caratti o Garatti, A., Stecklum, B., Garcia Lopez, R., et al.\ 2016, Nature Physics, 13, 276 

\bibitem[Carrasco-Gonz{\'a}lez et al.(2015)]{car15}
	Carrasco-Gonz{\'a}lez, C., Torrelles, J.~M., Cant{\'o}, J., et al.\ 2015, Science, 348, 114 

\bibitem[Davies et al.(2011)]{dav11}
	Davies, B., Hoare, M.~G., Lumsden, S.~L., et al.\ 2011, \mnras, 416, 972 

\bibitem[Ferland et al.(2013)]{fer13}
	Ferland, G.~J., Porter, R.~L., van Hoof, P.~A.~M., et al.\ 2013, RMxAA, 49, 137

\bibitem[Forbrich et al.(2016)]{for16}
	Forbrich, J., Rivilla, V.~M., Menten, K.~M., et al.\ 2016, \apj, 822, 93 

\bibitem[Garay et al.(1996)]{gar96}
	Garay, G., Ramirez, S., Rodriguez, L.~F., Curiel, S., \& Torrelles, J.~M.\ 1996, \apj, 459, 193 

\bibitem[Ghosh \& Lamb(1978)]{gho78}
	Ghosh, P., \& Lamb, F.~K.\ 1978, \apjl, 223, L83 

\bibitem[Goodman et al.(1993)]{goo93}
	Goodman, A.~A., Benson, P.~J., Fuller, G.~A., \& Myers, P.~C.\ 1993, \apj, 406, 528 

\bibitem[Guzm{\'a}n et al.(2012)]{guz12}
	Guzm{\'a}n, A.~E., Garay, G., Brooks, K.~J., \& Voronkov, M.~A.\ 2012, \apj, 753, 51

\bibitem[Hirota et al.(2015)]{hir15}
	Hirota, T., Kim, M.~K., Kurono, Y., \& Honma, M.\ 2015, \apj, 801, 82 

\bibitem[Hoare et al.(2007)]{hoa07}
	Hoare, M.~G., Kurtz, S.~E., Lizano, S., Keto, E., \& Hofner, P.\ 2007, Protostars and Planets V, 181 

\bibitem[Hosokawa \& Omukai(2009)]{hos09}
	Hosokawa, T., \& Omukai, K.\ 2009, \apj, 703, 1810

\bibitem[Hosokawa et al.(2010)]{hos10}
	Hosokawa, T., Yorke, H.~W., \& Omukai, K.\ 2010, \apj, 721, 478

\bibitem[Hunter et al.(2017)]{hun17}
	Hunter, T.~R., Brogan, C.~L., MacLeod, G., et al.\ 2017, \apjl, 837, L29 

\bibitem[Ilee et al.(2016)]{ill16}
	Ilee, J.~D., Cyganowski, C.~J., Nazari, P., et al.\ 2016, \mnras, 462, 4386 

\bibitem[Kim et al.(2008)]{kim08}
	Kim, M.~K., Hirota, T., Honma, M., et al.\ 2008, \pasj, 60, 991 

\bibitem[K\"onigl \& Pudritz(2000)]{kon00}
	K\"onigl, A., \& Pudritz, R.~E.\ 2000, Protostars and Planets IV, 759 

\bibitem[Krumholz et al.(2009)]{kru09}
	Krumholz, M.~R., Klein, R.~I., McKee, C.~F., Offner, S.~S.~R., \& Cunningham, A.~J.\ 2009, Science, 323, 754 
	
\bibitem[Kuiper et al.(2011)]{kui11}
	Kuiper, R., Klahr, H., Beuther, H., \& Henning, T.\ 2011, \apj, 732, 20 

\bibitem[Kurtz et al.(1994)]{kur94}
	Kurtz, S., Churchwell, E., \& Wood, D.~O.~S.\ 1994, \apjs, 91, 659 

\bibitem[Li et al.(2012)]{li12}
	Li, J., Wang, J., Gu, Q., Zhang, Z.-y., \& Zheng, X.\ 2012, \apj, 745, 47 

\bibitem[Matsushita et al.(2017)]{mat17}
	Matsushita, Y., Machida, M.~N., Sakurai, Y., \& Hosokawa, T.\ 2017, \mnras, 470, 1026 

\bibitem[McKee \& Tan(2003)]{mck03}
	McKee, C.~F., \& Tan, J.~C.\ 2003, \apj, 585, 850

\bibitem[McLaughlin \& Pudritz(1997)]{mcl97}
	McLaughlin, D.~E., \& Pudritz, R.~E.\ 1997, \apj, 476, 750 
	
\bibitem[Menten et al.(2007)]{men07}
	Menten, K.~M., Reid, M.~J., Forbrich, J., \& Brunthaler, A.\ 2007, \aap, 474, 515 

\bibitem[Meyer et al.(2017)]{mey17}
	Meyer, D.~M.-A., Vorobyov, E.~I., Kuiper, R., \& Kley, W.\ 2017, \mnras, 464, L90 

\bibitem[Mezger \& Henderson(1967)]{mez67}
	Mezger, P.~G., \& Henderson, A.~P.\ 1967, \apj, 147, 471 

\bibitem[Norman(2000)]{nor00}
	Norman, M.~L.\ 2000, Revista Mexicana de Astronomia y Astrofisica Conference Series, 9, 66

\bibitem[Ouyed \& Pudritz(1997)]{ouy97}
	Ouyed, R., \& Pudritz, R.~E.\ 1997, \apj, 482, 712 

\bibitem[Palau et al.(2013)]{pal13}
	Palau, A., Fuente, A., Girart, J.~M., et al.\ 2013, \apj, 762, 120 

\bibitem[Plambeck \& Wright(2016)]{pla16}
	Plambeck, R.~L., \& Wright, M.~C.~H.\ 2016, \apj, 833, 219 

\bibitem[Reynolds(1986)]{rey86}
	Reynolds, S.~P.\ 1986, \apj, 304, 713

\bibitem[Reid et al.(2007)]{rei07}
	Reid, M.~J., Menten, K.~M., Greenhill, L.~J., \& Chandler, C.~J.\ 2007, \apj, 664, 950 

\bibitem[Rodr{\'{\i}}guez et al.(2008)]{rod08}
	Rodr{\'{\i}}guez, L.~F., Moran, J.~M., Franco-Hern{\'a}ndez, R., et al.\ 2008, \aj, 135, 2370

\bibitem[Rosero et al.(2016)]{ros16}
	Rosero, V., Hofner, P., Claussen, M., et al.\ 2016, \apjs, 227, 25 

\bibitem[Shu(1977)]{shu77}
	Shu, F.~H.\ 1977, \apj, 214, 488 

\bibitem[Shu et al.(1987)]{shu87}
	Shu, F.~H., Adams, F.~C., \& Lizano, S.\ 1987, \araa, 25, 23 

\bibitem[Shu et al.(2000)]{shu00}
	Shu, F.~H., Najita, J.~R., Shang, H., \& Li, Z.-Y.\ 2000, Protostars and Planets IV, 789 

\bibitem[Smith et al.(2012)]{smi12}
	Smith, R.~J., Hosokawa, T., Omukai, K., Glover, S.~C.~O., \& Klessen, R.~S.\ 2012, \mnras, 424, 457 

\bibitem[Staff et al.(2015)]{sta15}
	Staff, J.~E., Koning, N., Ouyed, R., Thompson, A., \& Pudritz, R.~E.\ 2015, \mnras, 446, 3975 

\bibitem[Tan \& McKee(2003)]{tan03}
	in IAU Symp. 221, Star Formation at High Angular Resolution, ed. M. Burton, R. Jayawardhana, \& T. Bourke,
	arXiv:astro-ph/0309139

\bibitem[Tan et al.(2014)]{tan14}
	Tan, J.~C., Beltr{\'a}n, M.~T., Caselli, P., et al.\ 2014, Protostars and Planets VI, 149 
	
\bibitem[Tanaka et al.(2013)]{KT13}
	Tanaka, K.~E.~I., Nakamoto, T., \& Omukai, K.\ 2013, \apj, 773, 155

\bibitem[Tanaka et al.(2016)]{KT16}
	Tanaka, K.~E.~I., Tan, J.~C., \& Zhang, Y.\ 2016, \apj, 818, 52 (Paper I)

\bibitem[Tanaka et al.(2017)]{KT17}
	Tanaka, K.~E.~I., Tan, J.~C., \& Zhang, Y.\ 2017, \apj, 835, 32 

\bibitem[Tomida et al.(2017)]{tom17}
	Tomida, K., Machida, M.~N., Hosokawa, T., Sakurai, Y., \& Lin, C.~H.\ 2017, \apjl, 835, L11 

\bibitem[Vorobyov \& Basu(2015)]{vor15}
	Vorobyov, E.~I., \& Basu, S.\ 2015, \apj, 805, 115 

\bibitem[Wood \& Churchwell(1989)]{woo89}
	Wood, D.~O.~S., \& Churchwell, E.\ 1989, \apjs, 69, 831 

\bibitem[Zhang \& Tan(2011)]{zha11}
	Zhang, Y., \& Tan, J.~C.\ 2011, \apj, 733, 55 

\bibitem[Zhang \& Tan(2017)]{zha17}
	Zhang, Y., \& Tan, J.~C.\ 2017, arXiv:1708.08853 

\bibitem[Zhang et al.(2013a)]{zha13a}
	Zhang, Y., Tan, J.~C., De Buizer, J.~M., et al.\ 2013a, \apj, 767, 58

\bibitem[Zhang et al.(2013b)]{zha13}
	Zhang, Y., Tan, J.~C., \& McKee, C.~F.\ 2013b, \apj, 766, 86

\bibitem[Zhang et al.(2014)]{zha14}
	Zhang, Y., Tan, J.~C., \& Hosokawa, T.\ 2014, \apj, 788, 166

\end{thebibliography}
\end{document}